\documentstyle[a4,12pt]{article}

\newcommand{\bea}{\begin{eqnarray}}
\newcommand{\eea}{\end{eqnarray}}
\newcommand{\be}{\begin{equation}}
\newcommand{\ee}{\end{equation}}
\newcommand{\pkt}{\; .}
\newcommand{\kma}{\; ,}
\newcommand{\eqn}[1]{(\ref{#1})}
\newcommand{\tdot}[1]{\stackrel{\dots}{#1}}

\newcommand{\re}{{\rm Re}}

\newcommand{\abo}{\partial^\mu}
\newcommand{\abu}{\partial_\mu}

\newcommand{\om}{\omega_{k0}}
\newcommand{\intk}{\int\!\frac{{d^3}k }{(2\pi)^32\om}\,}
\newcommand{\intko}{\int\!\frac{{d^3}k }{(2\pi)^3\,}}

\newcommand{\calm}{{\cal M}}
\newcommand{\delm}{\delta m}
\newcommand{\dellam}{\delta \lambda}
\newcommand{\calv}{{\cal V}}
\newcommand{\calff}{{\cal F}_{\rm fin}}
\newcommand{\calfft}{\tilde\calff}
\newcommand{\imone}{I_{-1}}
\newcommand{\imthree}{I_{-3}}
\newcommand{\cc}{{\cal C}}

\newcommand{\calf}{{\cal F}}
\newcommand{\calc}{{\cal C}}

\newcommand{\cale}{{\cal E}}

\newcommand{\bfk}{{\bf k}}
\newcommand{\bfx}{{\bf x}}

\begin{document}
\begin{titlepage}
\begin{flushright}
DO-TH-00/07\\
March 2000
\end{flushright}

\vspace{20mm}
\begin{center}
{\Large \bf Nonequilibrium evolution and symmetry structure of the large-$N$
$\Phi^4$ model at finite temperature  }

\vspace{10mm}

{\large  J\"urgen Baacke\footnote{
e-mail:~baacke@physik.uni-dortmund.de} and Katrin Heitmann
\footnote{e-mail:~heitmann@hal1.physik.uni-dortmund.de} } \\
{\large Institut f\"ur Physik, Universit\"at Dortmund} \\
{\large D - 44221 Dortmund , Germany}
\\
\vspace{8mm}
\bf{Abstract}
\end{center}
We consider the large-$N$
$\Phi^4$ theory with spontaneously broken symmetry at finite temperature.
We study, in the large-N limit, quantum states which are characterized
by a time dependent, spatially homogeneous expectation value of one
of the field components, $\phi_N(t)$, and by quantum fluctuations
of the other $N-1$ components, that evolve in the background of the 
classical field. Investigating  such systems out of equilibrium
has recently been shown to display several interesting features. 
We extend here this type of investigations to finite temperature
systems. Essentially the novel features observed at $T=0$ carry over
to finite temperature. This is not unexpected, as the main 
mechanisms that determine the
late-time behavior remain the same. We extend two 
empirical - presumably exact - relations for the late-time
behavior to finite temperature
and use them to define the boundaries between the region of different
asymptotic regimes. This results in a phase diagram with the temperature
and the initial value of the classical field as parameters, the phases
being characterized by spontaneous symmetry breaking resp. symmetry
restoration. The time evolution is computed numerically and
agrees very well with the expectations. 
\end{titlepage}

\section{Introduction}
\setcounter{equation}{0}

The investigation of the $O(N)$ vector model at large $N$ has a
long-standing history in quantum field theory
\cite{CoJaPo,DoJa,BarMosh}.  One of the main aspects
was the question of symmetry restoration at high temperature that
for some time was controversial.
The dynamical exploration of a special class of nonequilibrium properties
has been developed only recently \cite{BdVHLS,CHKM,BDdVHS,BdVHS}. 

The out-of equilibrium configuration
that has been studied mainly is characterized by an initial
state in which one of the components has a 
spatially homogeneous classical expectation value
$\phi(t)$. This implies that the other $N-1$ components
$\psi_i(\bfx,t), i=1\dots N-1$ have a mass that is
different from the mass in the ground state. This means that their
initial state is related to the Fock space
vacuum state by a Bogoliubov transformation.
The evolution of the system is governed by the classical equation
of motion for the field $\phi(t)$ and by the mode equations 
for the quantum fields $\psi(\bfx,t)$. The expectation
value $\langle \psi(\bfx,t)\psi(\bfx,t)\rangle$ appears in both
equations of motion, this constitutes the quantum back reaction.
In the one-loop approximation, in contrast to the large-$N$ 
approximation, this quantum
back reaction only appears in the classical equation of motion.
This leads to decisive differences in the late time behavior.

We have previously \cite{BHPlnft} carried out such dynamical 
computations for the $O(N)$ vector 
model in the limit of large $N$  at finite temperature
for the case of unbroken symmetry, i.e., with a positive mass term.
Here we will consider the case of spontaneously broken symmetry.
In this case, at low temperatures the fields $\psi_i(\bfx,t)$
will be the Goldstone modes. This is the case for the ground state at
$T=0$ and at finite temperature; for
nonequilibrium initial states  these modes become massless
when the system settles to a stationary state at late times.
Symmetry restoration happens at high temperature {\em and} at large
values of the initial field $\phi(0)$; then at late times
these modes stay massive while the classical field vanishes, and thereby
the spontaneous symmetry breaking disappears.

Our investigation, as well as the analogous ones at $T=0$, are limited
to fields, masses (as solutions of the gap equation) and temperatures
much smaller than the scale of the Landau ghost 
$m_x=m_1 \exp(8\pi^2/\lambda)$, where $m_1$ is a renormalization scale,
taken  of order $\sqrt{\lambda} v$.
So the question of symmetry
non-restoration at ``really'' high temperatures \cite{BarMosh} 
will not be addressed here.

The plan of the paper is as follows:
in section 2 we introduce the model and set up the equations 
governing the nonequilibrium evolution.
In section 3 we discuss the renormalization of the
equations of motion and of the energy-momentum tensor,
some details are referred to Appendix A. 
In section 4 we discuss the phase structure
of the system as a function of temperature and initial conditions.
In section 5 we present the results of the numerical
computations. Some conclusions are drawn in section 6.


\section{Formulation of the model}
\setcounter{equation}{0}
We consider the $O(N)$ vector model with  the
Lagrangian 
\be
\label{lagrange}
{\cal L}={\displaystyle{\frac 1 2}}\abu \phi^i\abo\phi^i
-\frac{\lambda}{4N}(\phi^i\phi^i-N v^2)^2
\ee
where $\phi^i, i=1,..,N$ are $N$ real scalar fields.
The nonequilibrium state of the system is characterized
by a classical expectation value which we take in 
the direction of $\phi_N$. We split the field
into its expectation value $\phi$ and the quantum fluctuations
$\psi$ via
\begin{equation}
\label{erw}
\phi^i(\bfx,t)=\delta^i_N \sqrt{N} \phi(t)+\psi^i(\bfx,t)\pkt
\end{equation}
In the large-$N$ limit one neglects, in the Lagrangian,
all terms which are not of order $N$. In particular
terms containing the fluctuation $\psi_N$ of
the component $\phi_N$ are at most of order $\sqrt{N}$ and are
dropped, therefore. 
The fluctuations of the other components are identical, 
their summation produces
factors $N-1=N(1+O(1/N))$. In the broken symmetry case
these are the Goldstone modes.  Identifying all the fields
$\psi_1,..\psi_{N-1}$
as $\psi$ the leading order term in the Lagrangian then takes the form
\be \label{lnl1}
{\cal L}=N\left({\cal L}_{\phi}+{\cal L}_{\psi} + {\cal L}_I\right)\kma
\ee
with
\bea \label{lnl2}
{\cal L}_{\phi}&=&
\frac{1}{2}
\partial_\mu\phi\partial^\mu\phi-
\frac{\lambda}{4}\left(\phi^2-v^2\right)^2 \; ,\\
{\cal L}_{\psi}&=&\frac{1}{2}
\partial_\mu\psi\partial^\mu\psi+\frac{\lambda}{2}v^2\psi^2
+\frac{\lambda}{4}(\psi^2)^2 \kma \\
{\cal L}_{\rm I}&=&
-\frac{\lambda}{2}\psi^2\phi^2 \; ,
\eea
where $\psi^2$ is to be identified with $\sum \psi^i \psi^i/N$.

We decompose the fluctuating field into momentum
eigenfunctions via
\be
\psi(\bfx,t)=\intk \left[ a_\bfk U_k(t) e^{i\bfk\bfx}
+a^\dagger_\bfk U^*_k(t) e^{-i\bfk\bfx}\right]\kma
\ee
with $\om=\sqrt{m_0^2+k^2}$. The mass $m_0$ will be specified below.
This field decomposition defines a vacuum state as being annihilated
by the operators $a_\bfk$.

The equations of motion for the field $\phi(t)$ 
and of the fluctuations $U_k(t)$ have been derived in this
formalism by
various authors \cite{Boyanovsky:1994,Cooper:1994,Cooper:1995}.

We include in the following the counterterms that we will need
later in order to write the renormalized equations. The
equation of motion for the field $\phi$ becomes
\be
\ddot{\phi}(t)+\delta m^2 \phi(t)-\lambda v^2 \phi(t)+
(\lambda+\dellam)\phi(t)
\left[\phi^2(t)+\calf (t,T)\right]=0\pkt
\ee
Here $\calf (t,T)$ is the divergent fluctuation integral; it is
given by the average of the fluctuation fields defined by the
initial density matrix. For a thermal initial state of quanta with
energy $\omega_{k0}=\sqrt{k^2+m_0^2}$ it is given by
\be
\calf(t,T)=\langle \psi^2 (\bfx,t)\rangle
=\intk \coth{\frac{\beta\om}{2}}{|U_k(t)|^2}\pkt
\ee
The mode functions satisfy the equation:
\be
\label{mode}
\left[\frac{d^2}{dt^2}+\omega_k^2(t)\right]U_k(t)=0\kma
\ee
and the initial conditions
\be
U_k(0)=1 \;\; ;\;\; \dot U_k (0)= - i \om \pkt
\ee
The time dependent frequency $ \omega_k (t)$ is given by
\be
\omega^2_k(t) = k^2 +\calm^2(t)
\ee
with the time dependent mass
\be \label{effm}
\calm^2(t)=-\lambda v^2+\delm^2+(\lambda+\delta\lambda)\left[\phi^2(t)+
 \calf(t)\right]
\pkt\ee
Using this definition the classical equation of motion
can be rewritten as
\be
\label{eqmotion}
\ddot \phi(t)+\calm^2(t)\phi(t) =0
\ee
which is the same equation as the one for $U_k(t)$ with $k=0$
(zero mode).
Of course the initial conditions are different and $\phi(t)$ is real.

As in our previous work we rewrite the mode equation in the form
\be
\left[\frac{d^2}{dt^2}+\omega_{k0}^2\right]
U_k(t)=-\calv(t)U_k(t)\kma
\ee
whereby we have defined the time-dependent potential
$\calv(t)={\cal M}^2(t)-{\cal M}^2(0)$; we further identify
$m_0=\calm (0)$ as the ``initial mass''.

The average of energy with respect to the initial density matrix
is given by \footnote{Note that twice the last term, with positive sign,
is included in the fluctuation energy, since $\omega^2_k(t)$
contains $\calf(t,T)$.}
\bea
\label{energy}
{\cal E}&=&\frac{1}{2}\dot{\phi}^2(t)+{\displaystyle {\frac 1 2}}
(-\lambda v^2 +\delta m^2)\phi^2(t)
+\frac{\lambda+\dellam}{4}\phi^4(t)+\delta\Lambda
\nonumber\\
&&+\intk \coth{\frac{\beta\om}{2}}
\left\{{\displaystyle \frac 1 2}|\dot{U}_k(t)|^2
+{\displaystyle \frac{1}{2}}\omega^2_k(t)|U_k(t)|^2\right\}
\\ \nonumber &&-
\frac{\lambda+\dellam}{4}\calf^2(t,T)
\pkt
\eea
It is easy to check, using the equations of motion
(\ref{eqmotion}) and (\ref{mode}), that the energy is conserved.
The energy density is the $00$ component of the energy-momentum tensor.
The average of the energy momentum tensor for our system is diagonal,
its space-space components define the pressure which is given by
\bea \label{pdef}
p &=& \phi^2(t)-\cale + \delta \xi\frac{d^2}{dt^2}
\left[\phi^2(t)+\calf(t,T)\right]\\ \nonumber
&&+\intk \coth{\frac{\beta\om}{2}} \left(\om^2+\frac{k^2}{3}\right)
|U_k(t)|^2 \pkt
\eea
$\delta \xi$ is the renormalization of the conformal coupling
term $\xi (g_{\mu\nu}\partial^2-\partial_\mu\partial_\nu)\phi^2$, which has 
been used for the improved energy momentum tensor \cite{CaCoJa}.


\section{The renormalized equation of motion}
\setcounter{equation}{0}
The expressions for the time-dependent mass $\calm^2(t)$,
the energy density $\cale(t)$ and the pressure are still undefined
as they involve divergent integrals over the fluctuations.
Our approach to regularization and renormalization has been
presented previously \cite{BHP1,BHPlnft}. 
It is based on expanding the fluctuations
$U_k(t)$ and subsequently the various integrals involving
these fluctuations with respect to the time-dependent
potential $\calv(t)$. As this procedure has been presented
elsewhere in detail we just give the outline, here.

The expansion of the fluctuations with respect to $\calv(t)$
is given in Appendix A.
We use this perturbative expansion in order 
to single out the divergent contributions in the 
fluctuation integral. One finds
\be \label{flucex}
\calf(t)=\imone(m_0,T)
-\imthree(m_0,T)
\left[\calm^2(t)-\calm^2(0)\right]+\calff(t,T)\kma
\ee
where the finite part of $\calf(t,T)$ can be written as
\bea
\calff(t,T)
 &=&\intko
\frac{1}{4\omega_{k0}^3}\int\limits^{t}_{0}\!
{d}t'\cos\left[2\om(t-t')\right ]\dot{\calv}(t')
\coth{\frac{\beta\om}{2}} \nonumber\\
&+&\intko\frac{1}{2\om}\left[2 {\rm 
Re}f_k^{\overline {(2)}}(t)
+|f_k^{\overline {(1)}}(t)|^2\right]\coth{\frac{\beta\om}{2}} \kma
\eea
and where the divergent integrals are defined as
\bea
\label{imone}
I_{-1}(m_0,T)&=&\intko\frac{1}{2\om}
\left(1+\frac{2}{e^{\beta\omega_0}-1}\right)
=I_{-1}(m_0)+\Sigma_{-1}(m_0,T)\kma\\
\label{imthree}
I_{-3}(m_0,T)&=&\intko\frac{1}{4\om^3}
\left(1
+\frac{2}{e^{\beta\omega_0}-1}\right) 
=I_{-3}(m_0)+\Sigma_{-3}(m_0,T)
\pkt
\eea
The integrals $I_{-k}(m_0)$ are the genuine divergences which appear
in the renormalization at $T=0$.
Their dimensionally regularized form 
is given by
\bea
I_{-3}(m_0)&=&\left\{\intko \frac{1}{4\om^3}
\right\}_{\rm{reg}}=
 \frac{ 1}
{16\pi^2}\left\{\frac{2}{\epsilon}
+\ln{\frac{4\pi\mu^2}{m_0^2}}-\gamma\right\}\kma
\\ \nonumber
I_{-1}(m_0) &=&\left\{\intko\frac{1}{2\om}\right\}_{\rm{reg}}
=-\frac{ m_0^2}
{16\pi^2}\left\{\frac{2}{\epsilon}+
\ln{\frac{4\pi\mu^2}{m_0^2}}-\gamma+1\right\}
\\ \label{im1} &=&
-m_0^2 \imthree(m_0)-\frac{m_0^2}{16\pi^2}\pkt
\eea
 The additional temperature dependent terms $\Sigma_{-k}(m_0,T)$
are finite. They are defined as
\bea
\Sigma_{-1}(m_0,T)&=&
\intko\frac{1}{\om\left(e^{\beta\om}-1\right)}\kma\\
\Sigma_{-3}(m_0,T)&=&
\intko\frac{1}{2\om^3\left(e^{\beta\om}-1\right)} \pkt
\eea
It is convenient to include these
finite terms into the definition of $\calff(t,T)$.
Then the time dependent mass takes the form
\begin{equation}
\label{mass}
{\cal M}^2(t)=\lambda(\phi^2-v^2)+\delta\lambda\phi^2+
\delta m^2+(\lambda+\delta\lambda)\left[
I_{-1}(m_0)-I_{-3}(m_0){\cal V}(t)+\calfft(t,T)\right]
\kma
\end{equation}
with
\begin{equation} \label{fft}
\calfft(t,T)=\Sigma_{-1}(m_0,T)-{\cal V}(t)
\Sigma_{-3}(m_0,T)+
{\cal F}_{\rm fin}(t,T)\pkt
\end{equation}
The time dependent mass (\ref{mass}) contains both renormalization
constants $\delm$ and $\delta\lambda$. Furthermore, its definition by this
equation is implicit, $\calm^2(t)$ also appears  on the right
hand side of  (\ref{mass}) in $\calv(t)$. 

We now have to fix the renormalization counterterms in such
a way that the relation between  the time dependent mass 
and $\phi(t)$ becomes finite.
An additional constraint derives from the requirement
that the  renormalization
counterterms should not depend on the initial condition, but only
on the parameters appearing in the Lagrangian, i.e.,
$\lambda$ and $v$ and renormalization conventions. 

  We first determine $\delta\lambda$ by considering
the difference
\bea
\calv(t)&=&\calm^2(t)-\calm^2(0) \\ \nonumber 
&=&(\lambda+\delta\lambda)\left[\phi^2(t)-\phi^2(0)-I_{-3}(m_0)
\calv(t)+\calfft(t,T)-\calfft(0,T)\right]
\pkt
\eea
The divergent parts depend on the initial mass $m_0$. We have to replace this 
by a renormalization scale independent of the initial conditions. 
In Ref. \cite{BHPlnft} we had chosen the scale $m$, where $m$ was
the mass parameter appearing in the Lagrangian. Here the analogous
mass squared would be $m^2=-\lambda v^2$ and so $m$ would be imaginary.
We therefore choose another scale $m_1$ which we do not specify here.
In the numerical computations we have used the physical Higgs mass
$m_1^2=m_H^2 =2 \lambda v^2$.     

We rewrite the implicit equation for $\calv(t)$ as
\bea  \label{calvimp}
\calv(t)\left[1+(\lambda+\delta\lambda)I_{-3}(m_1)\right]&=&
(\lambda+\delta\lambda)\left\{\phi^2(t)-\phi^2(0)
-\left[I_{-3}(m_0)-I_{-3}(m_1)\right]
\calv(t)\right.\nonumber\\&&\left.+\calfft(t,T)-
\calfft(0,T)\right\}\eea
and require 
\be 
\label{dellamcond}
\frac{\lambda+\delta\lambda}{1+(\lambda+\delta\lambda)\imthree(m_1)}
=\lambda\pkt
\ee
Solving with respect to $\delta\lambda$ we find
\be \label{dellam}
\delta\lambda= \frac{\lambda^2\imthree(m_1)}
{1-\lambda\imthree(m_1)}\pkt
\ee
Inserting this relation into (\ref{calvimp}) we find
\be  \label{calvexp}
\calv(t)=\lambda\calc
\left[\phi^2(t)-\phi^2(0)
+\calfft(t,T)-\calfft(0,T)\right]\pkt
\ee
with
\be
\calc=\frac{1}{1+\lambda \left[\imthree(m_0)-\imthree(m_1)\right]}
=\frac{1}{\displaystyle 1+\frac{\lambda}{16\pi^2}
\ln\left(\frac{m_1^2}{m_0^2}\right)}
\pkt
\ee
Eq. \eqn{calvexp} is a finite relation for the potential $\calv(t)$
since the difference $[\imthree(m_0)-\imthree(m_1)]$ is finite.
Going back to Eq. \eqn{fft} we realize that $\calfft$ on the
right hand side contains itself
a term proportional to $\calv(t)$. Taking account of this term we rewrite
$\calv(t)$ in terms of $\calff$ as
\be  \label{calvexpT}
\calv(t)=\lambda\calc_T
\left[\phi^2(t)-\phi^2(0)
+\calff(t,T)\right]
\ee
with
\be
\calc_T=\frac{1}{\displaystyle 1+\frac{\lambda}{16\pi^2}
\ln\left(\frac{m_1^2}{m_0^2}\right)+\lambda \Sigma_{-3}(m_0,T)}
\pkt
\ee
Recall that $\calff(t)$ is the mode integral of 
second order in $\calv(t)$ and vanishes at $t=0$.

We now go back to equation (\ref{mass}) which we take at
the initial time $t=0$:
\be\label{lNmass0}
m_0^2\equiv\calm^2(0)=\lambda[\phi^2(0)-v^2]+\delta\lambda\phi^2(0)+
\delm^2+(\lambda+\delta\lambda)
\left[\imone(m_0)+\calfft(0,T)\right]\pkt
\ee
This is an implicit relation between $m_0$ and $\phi(0)$ which, however,
contains still the infinite quantities $\delta\lambda$, $\delm$ and
$\imone(m_0)$. 
Using Eq. \eqn{im1} we can rewrite Eq. \eqn{lNmass0} as
\be \label{gap2}
m_0^2=\left(-\lambda v^2+\delm^2\right)+(\lambda+\delta\lambda)
\left[\phi^2(0)-m_0^2\imthree(m_0)-
\frac{m_0^2}{16\pi^2}+\calfft(0,T)\right]\pkt
\ee
As renormalization condition we require $m_0$ to vanish,
for temperature $T=0$,
at the minimum of the potential $\phi=v$, as it is the case
on the tree level. We note that $m_0^2=0$ is {\em not} the
curvature of the tree level potential at $\phi=v$ which is
$m_H^2 = 2\lambda v^2$. It is the mass of the fluctuations at $\phi=v$ in
the large-$N$ approximation. For $T=0$ we have
$\calfft(t=0,T=0)=\Sigma_{-1}(m_0,T=0)=0$. Setting $m_0=0$, $\phi(0)=v$
in the gap equation \eqn{gap2} we get immediately
\be
\delm^2=-\delta \lambda v^2 =-\frac{\lambda^2v^2I_{-3}(m_1)}
{1-\lambda I_{-3}(m_1)}\pkt
\ee
Inserting this into Eq. \eqn{gap2} we obtain
 the renormalized gap equation
\be \label{gap3}
m_0^2=\lambda\calc\left[\phi^2(0)-v^2-\frac{m_0^2}{16\pi^2}
+\Sigma_{-1}(m_0,T)\right]
\pkt
\ee
For the numerical computation it is easier to choose some $m_0^2 \geq 0$
and to use the gap equation solved for $\phi^2(0)$:
\be \label{gapex}
\phi^2(0)=\frac{m_0^2}{\lambda}+v^2+
\frac{m_0^2}{16\pi^2}\left(1+\ln\frac{m_1^2}{m_0^2}\right)-
\Sigma_{-1}(m_0,T)
\pkt
\ee
For $t> 0$ the renormalized relation for the
mass squared $\calm^2(t)$ we find, using Eqns.
\eqn{calvexp} and \eqn{gap3}, is
\be \label{massfin}
\calm^2(t)=m_0^2+\calv(t)=\lambda\cc
\left[\phi^2(t)-v^2-\frac{m_0^2}{16\pi^2}
+\calfft(t,T)\right]\pkt
\ee
Having thus obtained a finite relation between $\phi(t)$ and $\calm(t)$ the 
equations of motion for the classical field $\phi(t)$ and for the
modes $U_k(t)$ are well-defined and finite.

The way in which we have renormalized has made the cutoff disappear.
This was possible only to the extent that we could safely neglect
corrections of order $\epsilon$ in the evaluation of the 
divergent integrals. One way of achieving this is
to take the limit $\epsilon \to 0$. This implies
for the bare coupling $\lambda_0$
\be
\lambda_0 = \lim_{\epsilon \to 0}
\frac{\lambda}{\displaystyle 1-\frac{\lambda}{16\pi^2}
\frac{2}{\epsilon}} = 0^{-}
\kma\ee
so this is the case of ``negative bare coupling'' as discussed in
\cite{BarMosh}. One can leave the cutoff finite, however, as long
as the masses and momenta are much smaller than the scale of the
Landau ghost, $m_x=m_1^2 \exp(8\pi^2/\lambda)$. This will be case
here. This is not related to a pragmatic momentum cutoff that we 
apply to the convergent integrals of the finite part.

While we have found here the gap equation as a self-consistency condition, 
it can also be derived \cite{CHKM,BdVHS} from a potential
(free energy) which here takes the form
\bea \nonumber 
V(m_0^2,\Phi^2,T)&=&\frac{m_0^2}{2}\left\{
\phi^2 - v^2 -\frac{m_0^2}{2\lambda}+\frac{m_0^2}{32\pi^2}
\left[\ln\left(\frac{m_0^2}{m_1^2}\right)-\frac{3}{2}\right]\right\}
\\
&&+\int\frac{d^3k}{(2\pi)^3}\frac{1}{\beta}
\ln\left[1-\exp(-\beta\omega_0)\right]
\pkt\eea
The gap equation then follows from the condition
\be
\frac{\partial V(m_0^2,\phi^2,T)}{\partial m_0^2}=0
\pkt
\ee
It should be mentioned here that the gap equation has two solutions,
one of which lies above the scale of the Landau ghost,
$m_x=m_1 \exp(8\pi^2/\lambda^2)$. In the sense that we consider here
the model as giving rise to a low energy effective theory we discard
this high mass solution, and its discussion. The solution we consider is the 
low energy one which is of order $\sqrt{\lambda} v$ or $m_1$.

The energy density is given by
\bea
\label{energyren}
{\cal E}&=&\frac{1}{2}\dot{\phi}^2(t)+{\displaystyle {\frac 1 4}}
\left(\lambda+\delta\lambda\right)\left(\phi^2-v^2\right)^2+\delta\Lambda
\nonumber\\
&& +\cale_{\rm fl}(t,T)-
\frac{\lambda+\delta\lambda}{4}\calf^2(t,T)
\pkt
\eea
Here we have used already that $\delta m^2 = -\delta\lambda v^2$, and
part of the ``cosmological constant'' counterterm $\delta \Lambda$ 
is included in $\delta\lambda v^4/4$.
The fluctuation energy is given by
\be
\cale_{\rm fl}(t,T)=\intk
\coth{\frac{\beta\om}{2}} \left\{\frac 1 2|\dot{U}_k(t)|^2
+\frac{1}{2}\omega^2_k(t)|U_k(t)|^2\right\}
\pkt
\ee 
We again split off the temperature-dependent contribution via
\be
\cale_{\rm fl}(t,T)=\cale_{\rm fl}(t,0)+\Delta\cale_{\rm fl}(t,T)
\kma\ee
where the second term on the right hand side 
\be \label{eflT} 
\Delta\cale_{\rm fl}(t,T)=\intk
\frac{2}{e^{\beta\om}-1} \left\{ \frac 1 2|\dot{U}_k(t)|^2
+\frac{1}{2}\omega^2_k(t)|U_k(t)|^2\right\}
\kma\ee
is finite. The divergences of the first term are given 
 \cite{BHP1} by the decomposition
\be 
\cale_{\rm fl}(t,0)=I_1(m_0)+\frac{1}{2}\calv(t)I_{-1}(m_0)
-\frac{1}{4}\calv^2(t)I_{-3}(m_0)+
\cale_{\rm fl,fin}(t,0)
\ee
with 
\be \label{efl0}
\cale_{\rm fl,fin}(t,0)
=
\frac{1}{2}\intk \left\{\frac{1}{2}|\dot f^{\overline{(1)}}_k|^2
+\frac{\calv(t)}{2}\left[2\re f_k^{\overline{(1)}}
+|f^{\overline{(1)}}_k|^2\right]+\frac{\calv^2(t)}{8\om^2}\right\}
\pkt\ee
We denote the sum of $\cale_{\rm fl,fin}(t,0)$ and
$\Delta\cale_{\rm fl}(t,T)$
finite contributions as $\cale_{\rm fl,fin}(t,T)$. The expression for
the energy then takes the form
\bea \label{ediv2}
\cale &=&\frac{1}{2}\dot \phi^2 
+\frac{\lambda+\delta\lambda}{4}\left(\phi^2-v^2\right)^2
 +\cale_{\rm fl,fin}(t,T)
+I_1(m_0)+\frac{1}{2}\calv(t)I_{-1}(m_0)
-\frac{1}{4}\calv^2(t)I_{-3}(m_0)  \nonumber \\
&&-\frac{\lambda+\delta\lambda}{4}\calf^2(t,T) 
+\delta\Lambda \pkt
\eea
In addition to the divergences arising from $\cale_{\rm fl}(t,T)$
we have to take into consideration those of $\calf^2(t,T)$ which
we have analyzed above. If all divergences and the renormalization
constant $\delta \lambda$ are inserted, the expression turns out to
be finite, i.e., the remaining counterterm $\delta \Lambda$ is needed
only for a finite renormalization. We require the energy to vanish
at $T=0$ for
$\phi(t)\equiv v$, which implies $m_0=0$. Then $\delta \Lambda =0$.
There remains a finite constant dependent on the initial condition
\be
\Delta \Lambda=  \frac{m_0^4}{128\pi^2}\left(1+\frac{2\lambda\calc}
{16\pi^2}\right) 
\ee
and the energy is given by
\bea \label{ediv}
\cale &=&\frac{1}{2}\dot \phi^2
 +\frac{\lambda}{4}\calc(\phi^2-v^2)^2 
 +\frac{1}{2}\Delta m^2 (\phi^2-v^2)
\\ \nonumber&&+\cale_{\rm fl,fin}(t,T)
-\frac{\lambda}{4}\cc\calfft^2(t,T) + \Delta \Lambda
\pkt\eea
Here $\Delta m^2$ is given by
\be
\Delta m^2=-\lambda\calc\frac{m_0^2}{16\pi^2}
\pkt
\ee
We write the pressure in the form
\be
p=\dot\phi^2(t)-\cale+p_{\rm fl}(t,T)+
\delta \xi \frac{d^2}{dt^2}\left[\phi^2(t)+\calf(t,T)\right]\pkt
\ee
The renormalization does not differ form the case
of unbroken symmetry discussed in Ref. \cite{BHPlnft}
and is not presented again.
We find
\be
\delta\xi=\frac{\lambda x}{6(1-\lambda x)}=\frac{\lambda I_{-3}(m)}
{6(1-\lambda I_{-3}(m))}
\pkt
\ee
The final result for the renormalized pressure reads
\be
p=\dot\phi^2(t)-\cale+p_{\rm fl,fin}(t,T)
-\frac{m_0^4}{96\pi^2}-\frac{m_0^2}{48\pi^2}\calv(t)
-\frac{1}{96\pi^2}\left[\ln\left(\frac{m_1^2}{m_0^2}\right)+2
\right]\ddot \calv(t)
\ee
with
\bea
p_{\rm fl,fin}(t,0)&=&
\intk
\left\{\left(\om^2+\frac{\vec k^2}{3}\right)
\left[2\re f_k^{\overline{(2)}}(t)+|f_k^{\overline{(1)}}(t)|^2
\right]\right.\nonumber\\
&&
+\left(\frac{1}{6\om^2}-\frac{m_0^2}{24\om^4}\right)
\int\limits_0^{t}\!dt'\,
\cos{2\omega_k^0(t-t')}\tdot{\calv}(t')\nonumber\\
&&+\left(\frac{1}{12\om^2}+\frac{m_0^2}{24\om^4}\right)
\cos(2\om t)\ddot{\calv}(0)\\
&&+|\dot{f}_k^{\overline{(1)}}(t)|^2
-2\re\left[i\om\dot{f}_k^{\overline{(1)}}(t)
+i\om f_k^{\overline{(1)}}(t)
f_k^{\overline{(1)}*}(t)\right]
\biggr\}\; .
\eea


\section{Analysis of the gap equation and of the phase structure}
\setcounter{equation}{0}

The dynamical evolution of the nonequilibrium system depends on two 
parameters, the temperature $T$ and the initial
amplitude of the classical field $\phi(0)=\phi_0$
which in analogy with thermal equilibrium systems can be considered
as an external parameter. There are two regions from which we can start 
the system, which we will call regions II and III.
There is, in addition, one region into which the system
can evolve when one considers $\phi(t=\infty)=\phi_\infty$ and not
$\phi_0$ as the external parameter. We call it region I.
In this section we will characterize these regions and
describe the dynamical evolution as to be expected from the
analysis at $T=0$. This analysis is based on certain empirical
results \cite{BDdVHS,BdVHS} that, though unproven, seem to be 
at least almost
exact. We will generalize these results to finite temperature
in a plausible heuristic way, to be confirmed by the numerical
computations. We think that the way in which we generalize 
these results will give a further clue to understanding them.


\subsection{Region I, $m_0^ 2 < 0$}

The gap equation requires $m_0^2$ to be positive. The point where
$m_0^2=0$ marks an initial condition that leads to a solution
$\phi={\rm const.}$, if $\dot \phi_0=0$ as we will assume in the following.
For $T=0$ this stationary amplitude is $\phi=v$. For $T> 0$ we can easily 
find this amplitude as well. Indeed for $m_0 =0$ the integral  
$\Sigma_{-1}(m_0,T)$ is given by its value for massless quanta,
i.e.,
\be \label{sigmahigh}
\Sigma_{-1}(0,T)=\frac{T^2}{12}
\kma 
\ee
therefore
\be \label{boundary12}
\phi_1^2(T)=\phi_0^2|_{m_0=0} = v^2-\frac{T^2}{12}
\pkt 
\ee
For $\phi_0 < \phi_1(T)$ the gap equation has no
real solution $m_0$. The region below the boundary
\eqn{boundary12} is region I. 

If nevertheless one wants to start the system with $\phi_0$ in region I
one faces the problem that in this region the gap equation requires
$m_0^2$ to be negative. Then the low-momentum modes
with $k^2 < - m_0^2$ have imaginary
frequencies. So from an orthodox point of view (to which we adhere here)
the system cannot be quantized properly. One may avoid this problem
by redefining the dispersion relation for the initial
frequencies via $\om^2= k^2+|m_0^2|$
in this region. Of course at $t>0$ $\calm^2(t)$ will be negative
so the ``mass squared'' changes sign at $T=0$, a situation called ``quench''
in analogy by a similar transition form a stable to an unstable state
by a sudden drop of temperature or inversion of a magnet field.
On the other hand the amplitude $\phi(t)$ can reach
this region at late times, but then it is in a quantum state
different form the ones we use as initial states.


\subsection{Region II: $m_0^ 2 > 0, m_\infty^2=0$}

We now assume $\phi_0$ is started above the boundary 
value \eqn{boundary12}. If $\phi_0$ is not too large
the system may, at $t> 0$, enter a  region 
where $\calm^2(t) < 0$, i.e., region I. 
Then the quantum fluctuations with momenta
$k^2 < - \calm^2(t)$ will increase exponentially, signalling instability.
This causes $\calm^2(t)$ to increase so that it is driven back to
a value $\calm^2(t) > 0$. If the initial amplitude $\phi_0$
is sufficiently small
this forth-and-back reaction will lead $\calm^2(t)$ to stabilize at
$\calm^2_\infty =0$. So at late times $\calfft(t,T) $ is determined by
quantum modes $U_k(t)$ that  oscillate with time-independent frequencies
$\omega_\infty =k$, it becomes stationary as well and
will be positive.
Therefore $\phi(t)$ stabilizes at some value
\be
\phi^2_\infty=v^2-\calfft(\infty,T) < v^2 - \calfft(0,T)
\pkt
\ee 
This is entirely analogous to the behavior found at $T=0$
\cite{BdVHS}. We call the region of initial values $\phi_0$
leading to this late-time behavior region II.

The stabilization by back-reaction onto the fluctuations 
obtained in the large-$N$
approximation  is not present in the {\em one-loop approximation}. 
In this approximation, once  $\phi(t)$ dips into the unstable region 
$\phi(t) < v/\sqrt{3}$, the mass squared of the fluctuations
becomes negative and the low momentum
modes evolve exponentially.   The effective mass of the 
classical field increases exponentially as well, and continues
to do so, but the mass squared of the
fluctuations stays negative. The amplitude $\phi(t)$ is driven towards 
zero. Nevertheless the classical energy continues to increase, as the
field oscillates faster and faster, this energy being extracted from the
energy of the quantum fluctuations. Obviously this signals the
instability of the quantum vacuum, as already apparent from the fact
that the effective potential is complex in this region. We will illustrate
this by a numerical example, to be presented in the next section.

At $T=0$ the final value $\phi_\infty$ was found to be related to the initial
value $\phi_0$ by an empirical relation 
\be \label{julienne}
\phi^2_\infty=\sqrt{\phi_0^2(2 v^2 -\phi_0^2)} \hspace{5mm} T=0
\pkt
\ee
It is not obvious how to generalize this relation to finite temperature.
It was remarked in Ref. \cite{BdVHS} that the relation only depends
on the initial, purely classical, energy, which is given by
$E=\lambda(\phi^2-v^2)^2/4$. Obviously it satisfies the constraints that
$\phi^2_\infty=v^2$ if $\phi_0=v^2$, and that $\phi_\infty=0$ if
classically  the system can reach the maximum of the  potential;
this happens at $\phi_0^2=\phi_2^2(T=0)=2v^2$.
So Eq. \eqn{julienne} seems to be related to energy considerations.
We further observe that the classical turning point is at
$\bar\phi_0^2=2 v^2 -\phi_0^2$ so that one may write
Eq. \eqn{julienne} as the geometric mean
\be
\phi^2_\infty=\sqrt{\phi_0^2\bar\phi_0^2}
\pkt
\ee
This form turns indeed out to lead to the correct generalization
for finite temperature.

Obviously the relation is characterized by the motion at early times
when the quantum fluctuations have not yet evolved.
When discussing renormalization we have made an expansion with respect
to the ``potential'' $\calv(t)$ which vanishes at $t=0$. So the same
expansion can be used to study the early time behavior. 
In the energy the coefficients of the terms of first
and second order in $\calv$ have been absorbed into renormalization
constants. However, the {\em thermal} fluctuations are not absorbed
in this way and will add to the classical terms in an early time
expansion. These appear in the energy, see Eq.\eqn{ediv2}, via
\be
\Delta\cale_{\rm fl}(t,T)=\Sigma_1(m_0)
+\frac{1}{2}\calv(t)\Sigma_{-1}(m_0)
-\frac{1}{4}\calv^2(t)\Sigma_{-3}(m_0) + O(\calv^3)
\ee
as a part of $\cale_{\rm fl,fin}(t,T)$ and via
Eq. \eqn{fft} in $\calfft(t,T)$.
Taking these expansions into account
 the energy can be written in the form
\be
E\simeq \frac{\lambda}{4}\calc
 \left[a \phi^4 + \tilde a \phi_0^4+b \phi^2 +\tilde b 
\phi_0^2+c\phi^2 \phi_0^2\right] +{\rm const.}
\ee
up to terms of order $\calv^3$. We need the coefficients
\bea
a&=&1-\lambda\calc_T\Sigma_{-3}(m_0,T)\kma
\\
b&=&-2 \left[v^2-\Sigma_{-1}(m_0,T)\right]\kma
\\
c&=&\lambda\calc\calc_T\Sigma_{-3}(m_0,T)\pkt
\eea
The classical turning point is given by
\be
\bar\phi_0^2=-\frac{b+(a+c)\phi_0^2}{a}
=\frac{1}{1-\lambda\calc_T\Sigma_{-3}}
\left[2 v^2 -\Sigma_{-1}-(1+\lambda\calc_T\Sigma_{-3})\phi_0^2\right]
\kma \ee
so that we are led to suppose
\be \label{assumix}
\phi_\infty^2(T)=\sqrt{\frac{1}{1-\lambda\calc_T\Sigma_{-3}}}
\sqrt{\phi_0^2\left
[2 v^2 -2 \Sigma_{-1}-(1+\lambda\calc_T\Sigma_{-3})\phi_0^2\right]}
\pkt\ee
We find indeed (see below) that this relation is very well 
fulfilled numerically.
According to this formula the region II is limited by
the requirement that the expression in the square root be positive,
so the boundary between region II and the new region III is given by
\be \label{boundary23}
\phi_2^2=2\frac{v^2 -\Sigma_{-1}(m_0,T)}
{1+\lambda\calc_T\Sigma_{-3}(m_0,T)}
\pkt\ee
We note that the relation is implicit,
the value of $m_0$ that appears on the right hand
side is related to $\phi_2^2$ on the left hand side by the
gap equation. 


\subsection{Region III, $\phi_\infty=0$ and $\calm_\infty^2 > 0$}

If the value $\phi_0$ becomes larger than $\phi_2$
the stationary state with constant $\phi$ and vanishing
mass $\calm^2(t)$ is no longer attained, and
the system reaches another asymptotic regime where
$\calm^2(\infty)\neq 0$ whereas $\phi(t) \to 0$. 
This regime is similar to the one that describes the 
late time behavior for the
unbroken symmetry case. We call the region of initial values  $\phi_0$
that leads to such a behavior region III.

There are two phenomena that characterize the transition to
this region. On the one hand the stabilization of the system
is taken over by the phenomenon of parametric resonance.
On the other hand the system has enough energy so that
$\phi(t)$ can move over the maximum of the potential at
$\phi=0$, and indeed will oscillate around $\phi=0$.
Accordingly the threshold value of $\phi_0$ at which 
these two phenomena set in can be characterized by
two - a priori unrelated - criteria. Both rely on plausible
assumptions, which at $T=0$ lead to the same prediction for
the critical value of $\phi_0$.

The criterion based on the energy consideration has been
presented in the previous subsection, we now
describe the criterion supplied by the phenomenon of
parametric resonance.
For the case of unbroken symmetry 
it was found at zero \cite{BDdVHS} and
finite temperature \cite{BHPlnft}, that
the late time behavior is described by an empirical 
sum rule which relates $\calm_\infty^2$ to the initial
amplitude. For $T=0$  an analogous sum rule was
found to hold for the case of spontaneously broken symmetry as well
\cite{BdVHS}.
It is given by
\be \label{sumrule1}
\mu_\infty^2 = - 1 + \frac{\eta_0^2}{2}
\pkt 
\ee
Here $\mu$ and $\eta$ are normalized in such a way that
the classical equation of motion at early times, i.e., in the
parametric resonance regime without back reaction, reads
\be \label{etanorm}
\eta'' -\eta+\eta^3 =0
\kma
\ee
where the prime denotes a derivative with respect to
$\tau = \alpha t$ and where $\eta = \beta \phi$, also
$\mu=\calm/\alpha$. With
$\eta(\tau)$ a solution of Eq. \eqn{etanorm} the mode equation
becomes a Lam\'e equation. The sum rule implies \cite{BDdVHS},
that the frequencies $\omega^2(t)=\calm^2(t)+k^2$ 
are shifted outside the parametric resonance band of the
Lam\'e equation. Though there is no rigorous derivation
for the sum rule, it accordingly seems  related to the parametric
resonance phenomenon. 

As the shift of the frequencies outside the
parametric resonance region must have happened at the end of the
phase where the evolution of the system is described
by parametric resonance, we will again consider
the initial classical evolution. Again, in addition to the
classical terms we have to take into account the terms due to the
thermal fluctuations.
In terms of the parameters introduced in the
previous section the equation of motion
is given by
\bea
\ddot\phi+\lambda \calc a \phi^3+\frac{\lambda}{2}\calc
(b+c\phi_0^2) \phi=0
\pkt\eea
Comparing to the normalized equation \eqn{etanorm}
we determine the factors $\alpha$ and $\beta$ to be
\bea
\alpha&=&\sqrt{\frac{\lambda\calc}{2}}\sqrt{b+c\phi_0^2}
\kma \\
\beta &=&\sqrt{-\frac{2a}{b+c\phi_0^2}}
\kma \eea
so that the asymptotic mass
is given by
\bea \label{sumrule2}
\calm_\infty^2&=&\alpha^2(-1 + \frac{1}{2}\beta^2\phi^2_0)
\\ \nonumber
&=&
\lambda\calc \left\{-v^2 +\Sigma_{-1}(m_0,T)+
\frac{1}{2}\left [1+\lambda \calc_T\Sigma_{-3}(m_0,T)\right]\phi_0^2
\right\}
\pkt
\eea
Again $\phi_0$ and $m_0$ are related by the gap equation.
At the transition from region II to region III the
asymptotic mass vanishes. It is easily seen that this
criterion leads to an identical equation for the boundary,
i.e., Eq. \eqn{boundary23}.

The field amplitude decreases to zero at late times, in this regime.
So the symmetry is restored dynamically at high excitation
characterized by a high value of $\phi_0$.

At the critical temperature $T=\sqrt{12} v$ both
boundaries $\phi_1(T)$ and $\phi_2(T)$ become zero.
Above $T_C$ the behavior of the system is the
same as for region III, for all initial values
of $\phi_0$. While at the border between 
region I and II there was a lowest value for
$\phi_0$ for obtaining real solutions
of the gap equation,  now
there is a lowest value of $m_0$, the one for which $\phi_0=0$.
It is obtained by solving the gap equation
for $\phi_0=0$ and agrees with the  thermodynamical
equilibrium value $m_\beta$ at that temperature, as defined, e.g., 
in Eq. (3.38) of Ref. \cite{DoJa}.
Of course with $\phi_0=0$ the system remains static.

Having defined the three regions by the two boundaries
\eqn{boundary12} and \eqn{boundary23} we present, 
in Fig. 1, a phase diagram in the
$\phi_0^2 - T $ plane. Fig. 2 shows the phase diagram in the
$m_0^2 - T$ plane, displaying, above $T_C$,  the region
$m_0<m_\beta$ which is excluded as an initial condition. 
We have to stress that the boundary between regions
II and III relies on an empirical relation.

The symmetry restoration above a critical temperature
is expected naively.
However, if the temperature becomes nonperturbatively large,
$T\simeq \sqrt{12} m_1 \exp(8\pi^2/\lambda)$, the gap equation 
does not have solutions any longer. Then the free energy attains its maximum
at the boundary $m_0=0$ and the $O(N)$ symmetry is again broken
\cite{BarMosh}.
This phenomenon of ``symmetry non-restoration'', 
as well as the existence of the second solution
of the gap equation above $m_x= m_1 \exp(8\pi^2/\lambda)$
will not be discussed here, as it is not part of the low energy
effective theory.


\section{Numerical Results}
\setcounter{equation}{0}

We have discussed already in the previous section the type of
nonequilibrium behavior to be expected in the different regions
of phase space. The numerical results follow these expectations.
We have chosen generally the parameters $v=1$ and $\lambda=1$.
We present results for the various regions in the $T,\phi_0$ plane.
The critical temperature is $2\sqrt{3}=3.464$. We choose
the temperatures between $T=1$ and $4$, the latter one being
above the phase transition.
The numerical method has been described in
\cite{BHPlnft}. We just recall that all the integrals
computed numerically are finite, so cutting off the momentum 
integration at some reasonable value is unrelated to
cutoffs used for renormalization. 

We first consider initial conditions in region II. 
The expectation value of $\phi$,
shown in Fig. 3 becomes
constant and different from zero as $t \to \infty$.
This signals spontaneous breakdown of
the $O(N)$ symmetry. As displayed in Fig. 4 the mass $\calm^2(t)$ vanishes
as $t \to \infty$, as expected form the Goldstone theorem.
 The momentum distribution of the quantum fluctuations
peaks at $k=0$ as $|U_k(t)|^ 2 \propto k^{-2}$, leading to long-range
correlations, a phenomenon called ``dynamical Bose-Einstein condensation''
in Ref. \cite{BdVHS} and investigated further, for finite volume,
in Ref.\cite{DM1}. We show an example of the momentum
distribution in Fig. 5, but we have not studied the phenomenon in detail.
 
The relation between the asymptotic value as $t\to \infty$
for $\phi(t)$ and the initial amplitude $\phi_0$ is 
displayed in Figs. 6 to 8, for $T=1,2.5$ and $3$.
 We compare the data with our generalization \eqn{assumix} of the  empirical
formula \eqn{julienne} given in Ref. \cite{BdVHS}.
The data are obtained by averaging over the
second half of the time interval. The agreement is 
excellent, except at the phase boundary where the averaging
converges slowly.

As an illustration of the behavior of the system in the unstable
region in the one-loop approximation we show, in Fig. 9, the
evolution of the field amplitude, and, in Fig. 10, the 
exponential behavior of the fluctuation integral and of the
effective mass squared $\calm^2(t)$ of the classical field.

The behavior of the system in region III is displayed in Figs.11 and 
12. The amplitude $\phi(t)$ is seen to decrease to zero. The decrease
is powerlike, not exponential, a phenomenon called
anomalous relaxation in Ref. \cite{BDdVHS}. Fig. 12 shows the squared mass
$\calm^2(t)$ which is seen to converge to an asymptotic value
$\calm_\infty^2$.
The sum rule for this asymptotic value, Eq. \eqn{assumix}, is
compared to the data in Fig. 13 for $T=1.5, 2.5$ and $4$.
The agreement is again excellent.

We have not presented the results for the pressure and the ratio
of pressure and energy which varies between $0$ for a nonrelativistic[D[D[D[Distic
and $1/3$ for an ultrarelativistic ensemble. Here these are dominated, 
already at $T=1$, by the purely thermal contributions, so that the fluctuations
generated by the motion of the field $\phi(t)$ are relatively unimportant.


\section{Conclusions and Outlook}
\setcounter{equation}{0}

The dynamical exploration of the quantum states of the $O(N)$
$\lambda \Phi^4$ theory in the limit $N \to \infty$ has been extended here
to finite temperature. We have performed numerical simulations
with various initial fields $\phi_0=\phi(0)$ and initial masses
$m_0$ related by the gap equation, and for various temperatures
$T$. Depending on the initial conditions
we find, in analogy to computations at zero temperature 
\cite{BDdVHS,BdVHS}, final states with restored $O(N)$ symmetry and
final states for which the symmetry is
spontaneously broken.
The resulting phase diagrams resemble typical phase diagrams of
thermodynamical systems, with the temperature and an external 
variable as parameters. Instead of, e.g., the magnetic field
or the pressure we have here the initial value $\phi_0$ as external
parameter. While the initial states are thermal states, the final states
are not. 

We have generalized two empirical formulae, the relation between
the initial and asymptotic field amplitudes in region II, and
the formula for the asymptotic value of $\calm^ 2(t)$ in region III
to finite temperature, extending the plausibility arguments given in
\cite{BdVHS}. While we have not been able, either, to derive these
formulae, the way of generalizing them may give some clue
for such a derivation. Both relations are linked as they give
the same formula for the boundary between regions II and III, though the
arguments for their heuristic derivation are seemingly different. Furthermore,
it is clear that both of them are based on the early time behavior.
Obviously the fluctuations have to be included up to
order $\calv^2(t)$ in a perturbative expansion. 
At $T=0$ these 
terms are essentially absorbed into renormalization constants,
so that the purely classical behavior prevails.
One may also formulate the modifications at finite temperature
in terms of temperature dependent masses and couplings.
It is the r\^ole
of the large-$N$ quantum back reaction to transmit the
early time behavior into the late time one. 

Unfortunately there are many interesting models for which
the large-$N$ approach is not possible or not adequate. 
The one-loop approximation, on the other hand,
 can be applied in general. However, it shows features that
seem to make it obsolete for describing nonequilibrium phenomena,
especially for theories with spontaneous symmetry
breaking. As an illustration we have shown the typical behavior 
of a spontaneously broken
$\lambda\Phi^ 4$ model in the one-loop approximation. 
The system  does not
reach a stationary state at late times: the effective mass of the 
classical field diverges exponentially, while the effective 
mass of the quantum fluctuations is and stays negative. This 
is due to the  lack of the quantum back reaction onto the fluctuations.
The fact that one finds such a pathological behavior may, however,
indicate the correct physics and is not 
necessarily a consequence of an inadequate approximation.
It is known that the system is indeed unstable for spatially
constant {\em static} fields,
it is an instability with respect to formation of domains
\cite{WeiWu}. 
For {\em space dependent} fields 
like minimal bubble  configurations the one-loop approximation 
to the effective {\em action} does not display any unplausible features
\cite{KriLaSch,Baa,Sue}, though
the effective {\em potential} is complex in the unstable region. 
So it is not clear whether the
``taming'' of the instability introduced by the large-$N$ approximation
necessarily improves the understanding of the physics.

In this situation it is certainly very important to develop new 
approaches to the evolution of quantum systems  for 
theories with spontaneously broken symmetry \cite{DM2,MACDH}.  
 There are indications in a large-$N$ quantum mechanical
system \cite{MACDH} that the large-$N$ limit may be misleading, as
the next-to-leading corrections become large especially at late times.
It is not clear, however, what the impact of these results
on quantum field theory will be.  One of the problems
is that, in contrast to the large-$N$ and one loop approximations,
alternative wave functionals pose problems with renormalization
\cite{DM2}. This is not only a technical problem. It is connected 
(trivially) to the fact that the higher the dimension of space, the more
the ultraviolet behavior of the system will be important.

We think nevertheless, that a good understanding of the
leading order approximation may improve the understanding of the
corrections. That these become large at late times is not too
surprising, it is therefore even more important to realize 
(once more) that the late-time behavior is related to the 
early-time behavior, which will therefore set the initial 
conditions for other approximations as well. This should apply to the
phase structure as well.


\section*{Acknowledgments}
The authors thank
 D. Cormier, H. de Vega, and J. Salgado for  useful discussions.
J. B. thanks the Deutsche Forschungsgemeinschaft
for partial support under grant No Ba 703/6-1.
K.H. thanks the Graduiertenkolleg
``Erzeugung und Zerf\"alle von Elementarteilchen'' for partial
support.


\begin{appendix}
\section{Perturbative expansion}
\setcounter{equation}{0}
The mode functions  $U_k(t)$  with the initial
conditions introduced in section 2
satisfy the  integral equation
\begin{equation} 
U_k(t)=e^{-i\omega_{k0} t}+
\int\limits^{\infty}_{0}\! dt'
\Delta_{k,{\rm ret}}(t-t')\calv(t')U_k(t')\;,
\end{equation}
with
\begin{equation} 
\label{fvt}
\Delta_{k,{\rm ret}}(t-t')= -\frac{1}{\om}
\Theta(t-t')\sin\left(\om(t-t')\right) \; .
\end{equation}
We separate $U_k(t)$ into the trivial part corresponding to
the case $\calv(t)=0$ and a function $f_k(t)$ which represents the
reaction to the potential by making the ansatz
\begin{equation} 
\label{ansatz}
U_k(t)=e^{-i\om t}[1+f_k(t)] \; .
\end{equation}
$f_k(t)$ satisfies then the integral equation
\begin{equation} \label{finteq}
f_k(t)=\int\limits^{t}_{0}\!dt'\Delta_{k,{\rm ret}}
(t-t')\calv(t')[1+f_k(t')]e^{i\om (t-t')}\;,
\end{equation}
and an equivalent differential equation
\begin{equation} \label{fdiffeq}
\ddot f_k(t)-2i\om \dot f_k(t)=-\calv(t)[1+f_k(t)]\;,
\end{equation}
with the initial conditions $f_k(0)=\dot{f}_k(0)=0$.
We expand now $f_k(t)$ with respect to orders in $\calv(t)$
by writing
\begin{eqnarray}
\label{entwicklung}
f_k(t)&=& f_k^{(1)}(t)+f_k^{(2)}(t)+f_k^{(3)}(t) +\cdots \\
 &=& f_k^{(1)}(t)+f_k^{{\overline{(2)}}}(t)
\; ,\end{eqnarray}
where $f_k^{(n)}(t)$ is of n'th order in $\calv(t)$ and 
$f_k^{\overline{(n)}}(t)$
is the sum over all orders beginning with the n'th one:
\begin{equation} 
f_k^{\overline{(n)}}(t)=\sum_{l=n}^\infty f_k^{(l)}(t)
\; .\end{equation}
The $f_k^{(n)}$ are obtained by iterating the integral
equation (\ref{finteq}) or the differential equation
(\ref{fdiffeq}). The function $f_k^{\overline{(1)}}(t)$ is
identical to the function $f_k(t)$ itself which is obtained
by solving (\ref{fdiffeq}). The function
$f_k^{\overline{(2)}}(t)$ can again be obtained
by iteration via
\begin{equation} \label{f2inteq}
f_k^{\overline{(2)}}(t)=
\int\limits^{t}_{0}\!{ d}t'\Delta_{k,{\rm ret}}
(t-t')\calv(t')
f_k^{\overline{(1)}}(t')e^{i\om (t-t')} \;.
\end{equation}
The integral equations can be used
in order to derive the asymptotic behavior as
$\om\to \infty$ and to separate divergent and finite 
contributions. This has been described previously in extenso
\cite{BHP1}.
We illustrate the procedure by calculating the relevant leading terms 
for $f_k^{(1)}(t)$. We have
\begin{equation} 
f_k^{(1)}(t)=\frac{i}{2 \om}
\int\limits^{t}_{0}\!dt'
(\exp(2 i \om(t-t'))-1)\calv(t')  \; .
\end{equation}
Integrating by parts we obtain
\begin{equation} \label{f1exp}
f_k^{(1)}(\tau)=
-\frac{i}{2\om}\int\limits^{t}_{0}\!{ d}t'
\calv(t')-\frac{1}{4\om^2}\calv(t)
+\frac{1}{4\om^2}\int\limits^{t}_{0}\!{d}t'
\exp(2 i \om(t-t'))\dot\calv(t')\;,
\end{equation}
For the expansion of the fluctuation integral $\calf(t)$
we need the real part of $f^{(1)}_k$ for which we find
\begin{equation}  \label{realexp}
 {\rm Re}\;h^{(1)}_k(t) =-\frac{1}{4\om^2}\calv(t)
+ \frac{1}{4\om^2}\int\limits^{t}_{0}\!dt'
\cos(2\om(t-t'))\dot\calv(t')
\pkt
\end{equation}
The second term decreases at least as $\om^{-3}$.
In terms of the perturbative expansion for the 
functions $f_k$ we can the mode functions 
appearing in the fluctuation integral as
\begin{equation} 
|U_k|^2=1+2 {\rm Re }\, f_k^{\overline{(1)}}+|{f_k^{\overline{(1)}}}|^2\pkt
\; \end{equation}
Using Eq.  \eqn{realexp} the leading behavior of this expression is
\begin{eqnarray}
\label{reskalsum}
1+2{\rm Re }\,{f}_k^{\overline{(1)}}+
|f_k^{\overline{(1)}}|^2\simeq1-\frac{1}{2\om^2}\calv(t)
\; .
\end{eqnarray}
Similarly the integrand of the energy density and pressure
can be expanded \cite{BHP1}. As these are more divergent,
 the calculations require more
integrations by parts in order to single out
the leading powers in $\om$ and they become more involved.

\end{appendix}

\newpage


\section*{Figure Captions}
\noindent
{\bf Fig. 1:} Phase diagram in the $\phi_0^2-T$ plane. \\
{\bf Fig. 2:} Phase diagram in the $m_0^2-T$ plane. \\
{\bf Fig. 3:} Evolution of classical field in region II. \\
{\bf Fig. 4:} Evolution of ${\cal M}^2(t)$ in region II. \\
{\bf Fig. 5:} The momentum spectrum for $T=1$ at $t=75$ displaying
``dynamical Bose-Einstein condensation'', with a fit
$sin^2(kt)/k^2$.\\
{\bf Fig. 6:} Late time amplitude $\phi(\infty)$ vs. 
initial amplitude $\phi_0$ for $T=1$ (asteriscs), compared with
Eq. \eqn{assumix} (solid line). \\
{\bf Fig. 7:} The same as Fig. 3 for $T=2$. \\
{\bf Fig. 8:} The same as Fig. 3 for $T=3$. \\
{\bf Fig. 9:} Evolution of the classical field in the one-loop approximation. \\
{\bf Fig. 10:} The fluctuation integral (solid line) and ${\cal M}^2(t)$
(dashed line) in the one-loop approximation. \\ 
{\bf Fig. 11:} Evolution of the classical field in region III.\\
{\bf Fig. 12:} Evolution of ${\cal M}^2(t)$ in region III. \\
{\bf Fig. 13:} The asymptotic sum rule for $\calm^2(t)$. The data
for $T=1.5$ (diamonds), $T=2.5$ (asteriscs) and
$T=4$ (triangles) are compared to Eq. \eqn{sumrule2} (solid lines). \\

\end{document}